\documentclass[authoryear,5p]{elsarticle}

\usepackage{graphicx}
\usepackage{amssymb}

\begin{document}

\title{Geometric interpretation of phyllotaxis transition}

\author{Takuya Okabe}
\address{
Faculty of Engineering, Shizuoka University, 3-5-1 Johoku, 
Hamamatsu 432-8561, Japan
}
\ead{ttokabe@ipc.shizuoka.ac.jp}


\begin{abstract}

The original problem 
of phyllotaxis was focused on 
the regular arrangements of leaves on mature stems 
represented by common fractions such as 1/2, 1/3, 2/5, 3/8, 5/13, etc. 
The phyllotaxis fraction is not fixed for each plant but it may 
 undergo stepwise 
transitions during ontogeny,  
despite contrasting observation that
the arrangement of leaf primordia at shoot apical meristems 
changes continuously. 
No explanation has been given so far for the mechanism of 
the phyllotaxis transition, 
excepting suggestion resorting to genetic programs 
operating at some specific stages.
Here it is pointed out that 
varying length of the leaf trace 
acts as an important factor 
 to control the transition  
by analyzing Larson's diagram of 
the procambial system of young cottonwood plants.  
The transition is interpreted 
as a necessary consequence of geometric constraints
that the leaf traces 
cannot be fitted into a fractional pattern unless 
their length is shorter than 
the denominator times the internode.  


\end{abstract}


\maketitle
\section{INTRODUCTION}

\begin{figure}
\begin{center}
\includegraphics[width=0.45\textwidth]{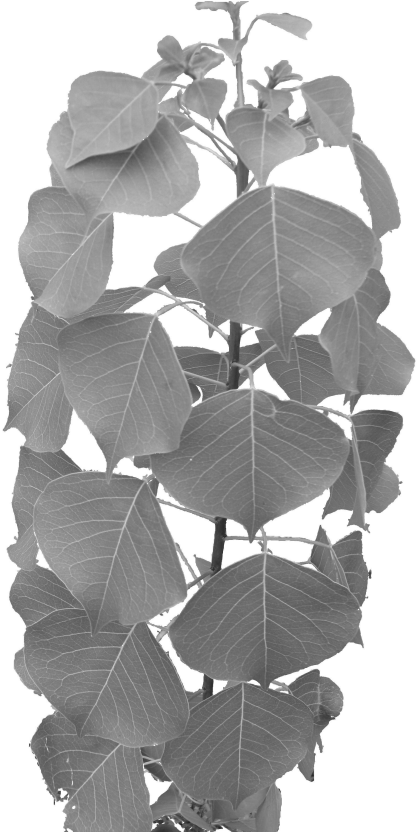}
\caption{A young poplar in a 3/8 phyllotaxis 
with eight vertical ranks of leaves.  
\label{fig0}}
\end{center}
\end{figure}

The spiral arrangement of leaves on a stem, phyllotaxis, 
is represented by the fraction of the circumference of the stem
traversed by the spiral in passing from one leaf to the next. 
Braun and Schimper discovered that 
leaves are lined up in ranks parallel to the stem so that 
the fraction is literally represented by a common
fraction (\cite{braun35}).  
In a 3/8 phyllotaxis, for instance, every eighth leaf comes over one below it after three turns of the
spiral,  so that eight straight ranks are visible along the stem (Fig.~\ref{fig0}). 
To this day, the following list of the most common fractions has been 
circulated in books and websites for non-specialists; 
1/2 for 
elm, lime and linden, 
1/3 for 
beech and hazel, 
2/5 for oak, 
cherry, apple, holly and plum, 
3/8 for poplar, rose and pear, 
5/13 for almond, etc. 
Such correspondence tables seem to have existed already 
in the middle of the nineteenth century (\cite{henfrey70}). 
Sometimes willow is listed in 5/13 (\cite{coxeter61})
and sometimes in 3/8 (\cite{adam06}). 
As a matter of fact, 
it had been remarked 
since early times 
that even an individual plant sometimes 
makes transitions between different fractions (\cite{braun35}).
Therefore, the phyllotaxis fraction is not a determined characteristic of each species. 
Most notably, \cite{larson80} 
has revealed the manner in which the vascular system is rearranged
through the phyllotaxis transition  
by mapping arrangement of the leaf traces, 
the portion of vascular bundles of leaves that resides in the
stem (Fig.~\ref{fig1}).  
By contrast, it has been commonly accepted that 
the arrangement of leaf primordia in the bud or at the shoot apical meristem
 does not conform to any fractional number; 
primordia do not appear as radial rows. 
The angular divergence between successive primordia stays 
close to a unique ``ideal'' angle of 137.5$^\circ$, 
which is the golden mean 0.3820  of 360$^\circ$ (\cite{church20,richards51}). 
The golden mean is a mathematical limit of the sequence of the above fractions.  
Much attention has been paid to mathematics of these numbers
and to mechanisms of leaf primordia formation by which the ideal angle
is achieved and regulated (\cite{abj97}). 
In contrast, 
the phyllotaxis transition has been left unexplained 
without attracting interest from researchers.
The fractional phyllotaxis and the phyllotaxis transition are two sides of the same problem. 
A key point noted in the present study 
is that 
phyllotaxis in the bud changes {\it continuously} 
 whereas phyllotaxis on the mature stem 
changes {\it stepwise}.  
%
%
Evidently,  the arrangement of primordia at the apical growing point
must be a determining factor of leaf arrangement on the mature stem. 
But what causes the transition from 2/5 to 3/8, for instance?  
There must be another factor. 
\cite{larson80} has suggested that the transition is programmed in the plant  
to occur at specific stages of ontogeny.
In another view, phyllotaxis of primordium formation 
and vascularization are controlled by some higher-level system (\cite{rhh93}). 
\cite{kuhlemier07} remarks that virtually nothing is known about the molecular
mechanisms that underlie the transitions between different
spiral systems (e.g. 3/8 to 5/13 patterns), except
that larger meristems seem to have higher Fibonacci numbers. 
The present paper aims at pointing out that this phenomenon is 
consistently explained by considering geometry of growing leaf traces.  
%
%
By means of a full quantitative analysis, which surely is not standard in this field of research,  
it is shown  
(1) that the positions of the phyllotaxis transition are located 
based on 
the angular positions at which leaf traces exit from the
vascular cylinder 
and (2) that the phyllotaxis transition is 
caused 
as a necessary consequence of change in size of leaf traces relative to
internode length. 

%
%

A caveat: this paper deals with the mechanism of the phyllotaxis transition in spiral systems. 
Although a common keyword of ``phyllotaxis''  may suggest, 
it should not be confused with 
mechanisms of phyllotactic primordium formation, 
for which considerable advances have been made over the last decade (\cite{kuhlemier07}). 
Models of the latter category deal with continuous changes of apical
meristems, while they are not concerned with 
the fractional expression of  a phyllotactic pattern.




%

%
%
%
%
%
%
%

\section{MATERIAL AND METHODS}

\begin{figure}
\begin{center}
\includegraphics[width=0.48\textwidth]{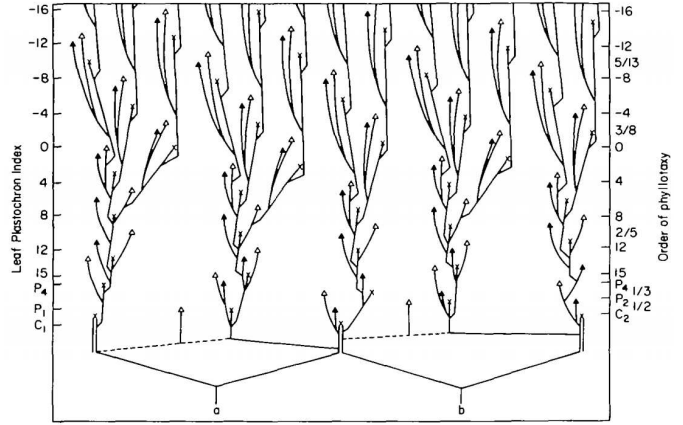}
\caption{
Diagram of the procambial system of a typical cottonwood plant 
compiled by \cite{larson80}. 
The vascular cylinder is displayed as if unrolled and laid flat. 
The ordinate is the Leaf Plastochron Index (LPI) for leaves, whereas 
the abscissa corresponds to the angular coordinate in a full turn about the stem axis. 
Each leaf is entered by three traces; a central ($\times$), 
right ($\blacktriangle$), 
and left ($\triangle$) traces.  
Phyllotaxis orders, 1/2, 1/3, 2/5, 3/8 and 5/13, are  indicated by the
 right vertical axis. 
\label{fig1}
}
\end{center}
\end{figure}

\begin{figure}[t]
\begin{center}
\includegraphics[width=0.48\textwidth]{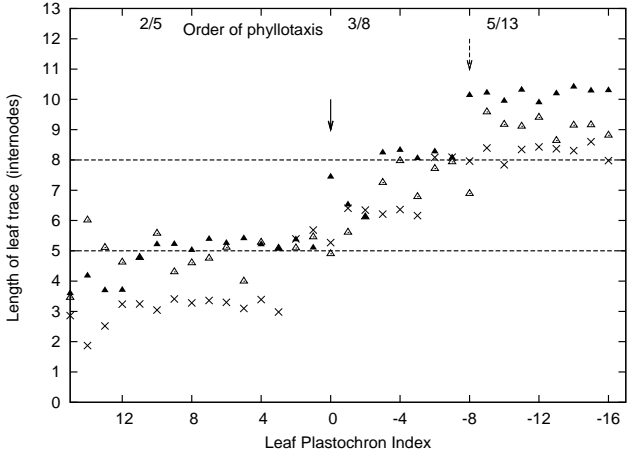}
\caption{
Vertical length of leaf traces in Fig.~\ref{fig1}
is plotted against the Leaf Plastochron Index. 
Orders of phyllotaxis, 2/5, 3/8 and 5/13, are denoted 
at the same position as indicated in Fig.~\ref{fig1}. 
See Fig.~\ref{fig1} for trace designations. 
The phyllotaxis transitions from 2/5 to 3/8  and 
from 3/8 to 5/13 are caused by 
the traces indicated by a solid arrow and a dashed arrow, respectively. 
Horizontal dashed lines at 5 and 8, separating the different phyllotaxis regimes, 
are drawn for reference sake (see Results and Discussion).  
\label{fig2}
}
\end{center}
\end{figure}

The analysis is based on a diagram 
of the procambial system of a cottonwood plant ({\it Populus deltoides}) 
reconstructed by \cite{larson80}, which is reproduced in Fig.~\ref{fig1}. 
The vascular cylinder is displayed as if unrolled and laid flat. 
The ordinate is the leaf plastochron index (LPI)  for numbering leaves. 
Each leaf has three traces: central, right and left traces exit the vascular cylinder 
at positions denoted by symbols $\times$, $\triangle$ and $\blacktriangle$,
respectively.  
For more details, see \cite{larson80} and references cited therein. 

By virtue of the convention to take LPI as the ordinate, 
the vertical scale of the diagram is normalized such that 
differences in height between two successive leaves, or internodes, 
are a unit of length in the vertical direction. 
Therefore, 
the vertical component of a line segment 
in Fig.~\ref{fig1} represents not its actual length  
but an effective length relative to the internode length, 
namely the length measured in internode units.  
%
%
Accordingly, vertical lengths of the leaf traces in internode units 
are directly evaluated by applying a digitizing ruler to the diagram.   
%
The lower end points of right and left traces are located without ambiguity, 
for they are connected to other types of traces. 
%
%
Lengths of central traces are fixed by decomposing the whole pattern
into clusters consisting of three adjacent traces, the right trace of leaf $n$, 
the central trace of leaf $n+2$  and 
the left trace of leaf $n-1$, where $n$ for LPI is an integer. 
For illustration purposes, the clusters 
from $n=0$ to 7 are shown in Fig.~\ref{fig4+}. 
The length of the leaf traces thus obtained 
is plotted against LPI in Fig.~\ref{fig2}, 
where reading errors are of the order of an internode at most. 
On the other hand, 
divergence angles are evaluated from the horizontal, angular coordinates
of the exit points of the leaf traces denoted by the symbols. 
Evaluated angles show rhythmic, systematic variations around the
``ideal'' angle, which are typically observed in a quantitative analysis (\cite{okabe12b}).  
%
The systematic variations, which are correlated with the angular positions, 
are suppressed apparently by deleting from the diagram a blank
rectangular strip of a narrow width along the left ordinate.
The width of the strip is determined so as to minimize the standard deviation of 
divergence angles for ten youngest central traces (LPI less than $-6$).  
Results for the divergence angle thus corrected are shown in Fig.~\ref{fig3}. 
The width ratio of the deleted strip is 0.023, 
whereby the standard deviation of the divergence angle 
is suppressed from 4.2$^\circ$ to 0.89$^\circ$. 
The correction is made for ease of understanding implications of Fig.~\ref{fig3}. 
It does not affect the results discussed below qualitatively. 



\section{RESULTS}

\begin{figure}[t]
\begin{center}
\includegraphics[width=0.49\textwidth]{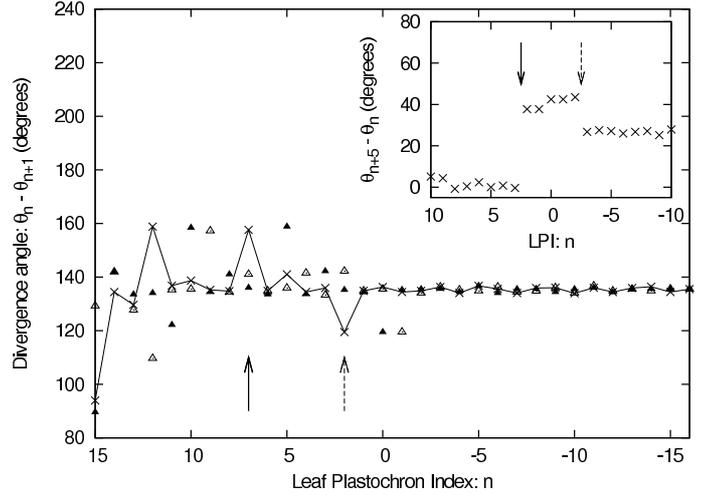} 
\caption{
Divergence angle between the traces at LPI $n$ and $n+1$ 
is plotted against LPI $n$. 
See Fig.~\ref{fig1} for trace designations. 
The inset shows the angle between the central traces at LPI $n$ and $n+5$:  
for ideal patterns of 2/5, 3/8 and 5/13 orders, 
it should be zero, 45 and 28 degrees, respectively.
The positions at which 
the phyllotaxis transitions from 2/5 to 3/8 and from 3/8 to 5/13 
occur are indicated by solid and dashed arrows, respectively. 
\label{fig3}
}
\end{center}
\end{figure}

Results in Fig.~\ref{fig2} indicate that 
 the three traces grow in length steadily all alike.  
This behavior is consistent with other quantities reported by
\cite{larson80}.
According to Fig.~\ref{fig2}, 
three traces at LPI 2 have almost the same length of about five internodes. 
This means that the traces at LPI 2 extend down to the height of LPI 7 (see Fig.~\ref{fig1}). 
%
As indicated by the right ordinate of Fig.~\ref{fig1}, 
the phyllotaxis order fraction changes from 2/5 through 3/8 to 5/13 
as we climb up the stem, or as LPI decreases. 
The order fractions are indicated at the top of Fig.~\ref{fig2} 
at the same LPI coordinates as in Fig.~\ref{fig1} by \cite{larson80}.  
Horizontal dashed lines at five and eight internodes in Fig.~\ref{fig2}
are drawn to separate regimes of different phyllotaxis orders (see below). 

Divergence angle between the leaf traces at LPI $n$ and $n+1$ 
is plotted against LPI $n$ in Fig.~\ref{fig3}. 
The inset shows the angle between the central traces at LPI $n$ and
$n+5$, that is, the net angle of inclination of 5-parastichies. 
The angle should become zero, 45 and 28 degrees for 2/5, 3/8 and
5/13 ideal patterns, respectively; 
for instance, 
the ideal angle for 3/8 is $360\times 3/8 \times 5 = 360\times 2 -45$,
which is congruent to $-45$. 
The inset of Fig.~\ref{fig3} clearly indicates stepwise transitions
between the three distinct fraction regimes. 
%
%
Thus, the positions of the phyllotaxis transition are located based on
the exit points, or the bases, of the leaf traces, i.e., 
without inspecting internal changes in the vascular structure. 
This crucial property for us is brought to light 
 probably because the subject plants are grown under
controlled uniform conditions (\cite{larson80}. 

%
%

The cause of the transition is traced back by 
close inspection based on the quantitative results. 
%
%
The transition from 3/8 to 5/13, 
indicated by a dashed arrow in the inset of Fig.~\ref{fig3}, 
is caused by an irregular decrease of divergence angle at LPI 2, 
which therefore is also indicated by a dashed arrow in the main figure. 
Similarly, a solid arrow in the inset indicates the transition from 2/5
to 3/8, which is ascribed to an increase of divergence angle at LPI 7, 
a solid arrow 
in the main figure. 

In connection with 
 the latter transition,  
it is worth a remark that 
divergence angle in the 2/5 phyllotaxis regime 
is not held at an ideal constant value of 144$^\circ$ $(360\times 2/5=144)$. 
This means that five ranks (orthostichies) of a real 2/5 pattern is not equally spaced. 
According to Fig.~\ref{fig3}, 
two full turns ($360^\circ \times 2$) 
of the 2/5 pattern is divided roughly into unequal parts of 
$140^\circ\times 4+160^\circ\times 1$, instead of 
a regular spacing with $144^\circ \times 5$. 
The figure shows that the irregular shift at LPI 7 is shared with LPI 12. 
This is just as expected for the 2/5 arrangement $(7+5=12)$. 
Similarly,  
a cycle of $360^\circ \times 3$ of a 3/8 pattern  is divided into 
$137^\circ\times 7+120^\circ\times 1$, 
as indicated by the dashed arrow.  
%
Thus, 
the quantitative analysis reveals that 
divergence angle {\it on a stem} is a secondary property, 
as it is very unlikely that the exceptional angles of $160^\circ$ and
$120^\circ$ are intrinsic to the plant. 
%
%
%
Accordingly, the pattern of Fig.~\ref{fig1} 
is to be viewed as a result of secondary processes. 

\begin{figure}[t]
\begin{center}
\includegraphics[width=0.49\textwidth]{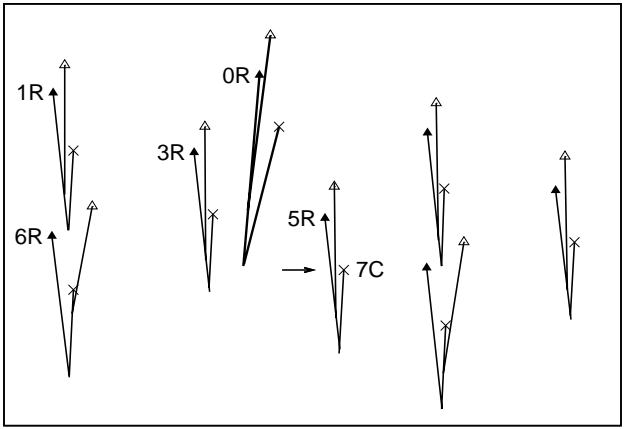} 
\caption{
A schematic excerpt from Fig.~\ref{fig1} 
near the transition from 2/5 to 3/8. 
The right trace at LPI 5 is denoted as 5R, and 7C signifies the central trace at LPI 7. 
According to Fig.~\ref{fig3}, 
the traces 5R and 7C 
are shifted slightly to the right  
as compared to the preceding (lower) traces. 
This figure shows that the shift (solid arrow) is caused 
by a longer trace 0R intervening between 3R and 5R, 
thereby 
the transition 
is initiated.  
If the length of 0R were shorter than five internodes, 
0R should have been aligned with 
5R to keep the 2/5 order, 
as the preceding 1R is with 6R.  
\label{fig4+}
}
\end{center}
\end{figure}

The final step is to identify the cause of the irregular shift in divergence angle.  
For brevity, let the central ($\times$), right ($\blacktriangle$) and 
left ($\triangle$) trace of LPI $n$ be denoted as 
$n$C, $n$R and $n$L, respectively. 
At the transition from 2/5 to 3/8, 
the irregular shift of 7C (solid arrow)
is accompanied by 5R on the right side (Fig.~\ref{fig3}).   
%
%
%
Inspection of Fig.~\ref{fig1} reveals that 
this collective shift is caused as a result of 0R intervening between 3R
and 5R. 
This is illustrated in Fig.~\ref{fig4+}, a schematic excerpt from Fig.~\ref{fig1}.  
For this reason, the trace 0R is indicated by a solid arrow in Fig.~\ref{fig2} 
as the very cause of the phyllotaxis transition from 2/5 to 3/8.   
A horizontal line at five is drawn in Fig.~\ref{fig2}
to indicate that 0R interferes with 5R if only the former length exceeds 
$5-0=5$ internodes (cf. Fig.~\ref{fig4+}).  
Indeed, the filled triangle at the solid arrow in Fig.~\ref{fig2} 
lies well above the horizontal line at five internodes. 
%
Similarly, the trace causing the transition from 3/8 to 5/13 is indicated 
by a dashed arrow in Fig.~\ref{fig2}, where the upper horizontal line at eight is 
drawn as a threshold length for the transition. 

\section{DISCUSSION}

\begin{figure}[t]
\begin{center}
\includegraphics[width=0.49\textwidth]{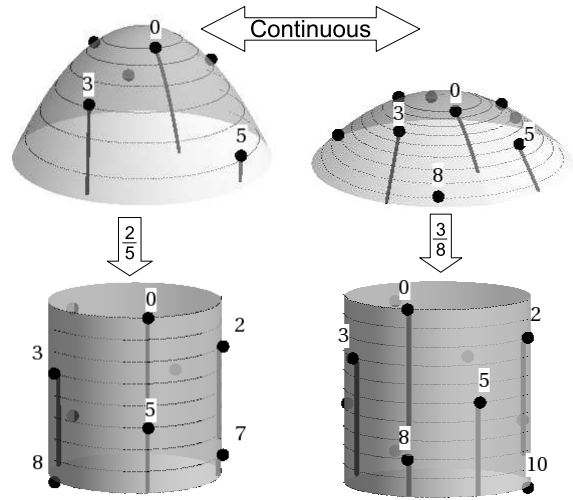} 
\caption{
Distinct patterns 
of 2/5 and 3/8 orders
on the mature stem cylinders (bottom) 
result from similar arrangements of leaf primordia 
at the shoot apical meristems (top).
Each dot represents a leaf node,
through which a dotted line is drawn to demarcate internodes.
Leaf traces (line segments) 
are aligned in a 2/5 phyllotaxis pattern 
if their length is greater than three and less than five internodes (left), 
whereas a 3/8 phyllotaxis results if it is greater than five and less than eight internodes (right). 
\label{fig4}
}
\end{center}
\end{figure}

%
%
%
%
%
%
%
%

The trace length represented in internode units is 
an important geometric factor 
as it imposes 
constraints on possible fractional
patterns to be realized (\cite{okabe11,okabe12}). 
The geometric effect invoked in interpreting the above results 
is schematically illustrated in Fig.~\ref{fig4}. 
During growth of the shoot apex, the precursor of the vascular system  
associated with a leaf primordium develops to become the leaf trace (\cite{rhh93}).
Both lengths of the leaf trace and internode change continuously during growth. 
In this view, 
the phyllotaxis order to be realized on the mature stem is determined
depending on the trace length represented in internode units.
%
%
Top two patterns in Fig.~\ref{fig4} represent the initial arrangements 
of the most common case of 137.5$^\circ$ angular divergence,   
whereas their growth rates in the radial direction are different, 
i.e.,   the two patterns are characterized 
with different plastochron ratios (\cite{richards51}). 
At this point, the difference is not a qualitative but a quantitative one. 
Leaf traces (line segments) in the left pattern traverse about four internodes, 
while those in the right pattern traverse about six internodes. 
As the figure shows,  a qualitative difference is brought about 
through the quantitative difference in the trace length: 
geometric constraints tend to achieve 
either a 2/5 or 3/8 phyllotaxis on the mature stem depending on 
whether the trace length is shorter or longer than five internodes. 
Experimentally, 
the difference would be judged on 
whether leaf trace 0 is identified or not 
in the cross section at the level of node 5. 
In a similar manner, 
the threshold value of the trace length for the transition from 3/8 to 5/13 order 
is eight internodes, the denominator of the lower order fraction. 
These are indicated by horizontal dashed lines in Fig.~\ref{fig2}, as already remarked. 
In terms of the threshold length, 
the phyllotaxis transition is interpreted consistently 
without resort to operations of elaborate mechanisms like genetic
programs; 
 the phyllotaxis transitions are caused because 
changing length of the leaf trace 
{\it happens to} cross the threshold values of five and eight internodes.
As illustrated in Fig.~\ref{fig4}, 
this view provides a consistent 
explanation of empirical observation that 
large meristems result in arrangements of higher order fractions. 
Unequal distribution of divergence angle noted in Fig.~\ref{fig3} 
is circumstantial evidence of a unique intrinsic angle close
to 137.5$^\circ$ and secondary distortions therefrom.


%
%

%

%
%
%

Fibonacci numbers
5 and 8 enter as the threshold values  
because leaf 0 
appears close to 
leaves 5 and 8 (Fig.~\ref{fig4}).   
This in turn is a mathematical consequence of 
the golden-mean angle at the apex.
Mathematically, 
any number can be approximated by a common fraction with any assigned
degree of accuracy. 
The greater the denominator, the better the approximation. 
A fast-converging sequence of approximate fractions 
is uniquely determined for a given number (\cite{hw79}). 
For the golden mean 0.3820, it is 1/2, 1/3, 2/5, 3/8, 5/13, etc., 
the main sequence of phyllotaxis.  
The denominators of these approximate fractions 
comprise the index differences of nearby leaves, namely 
the threshold values for the trace length. 
Thus, the geometric interpretation predicts 
a correlation between the fraction index of the phyllotaxis order and 
the trace length represented in internode units. 
Indeed, it has been remarked as a general rule; 
the higher phyllotaxis order is associated with the longer leaf trace (\cite{girolami53,esau65}). 
To conclude, the plant in its maturity, as it were, 
achieves 
rational approximations to a divergence angle at the apex 
in conformity with the leaf-trace length in internode units.

Mathematically, 
the golden mean is the worst ``approximable'' real number 
in the sense that the sequence of the approximate fractions converges most badly. 
In this connection, it has been commonly mentioned, despite objections, 
 that the golden-mean divergence is advantageous because 
leaves are distributed most evenly to sunlight.   
When viewed in the context of this study, 
the golden-mean divergence distributes the leaf traces most efficiently 
to coordinate the vascular system.
%
%
Although 
``phyllotaxis'' is the arrangement of leaves {on} a stem 
according to dictionaries,  
 the geometric view on the leaf-trace organization,  
the arrangement of leaves {in} a stem,  
may shed a new light on the long-standing problem of phyllotaxis in vascular plants.

%
%
%
%


\bibliography{okabe}

\end{document}